# Prediction of grain boundary structure and energy by machine learning

Shin Kiyohara,[1] Tomohiro Miyata[1] and Teruyasu Mizoguchi[1]*

[1]Institute of Industrial Science, University of Tokyo, 4-6-1, Komaba, Meguro, 153-8505 Tokyo, Japan

**Abstract**

Grain boundaries dramatically affect the properties of polycrystalline materials because of differences in atomic configuration. To fully understand the relationship between grain boundaries and materials properties, systematic studies of the grain boundary atomic structure are crucial. However, such studies are limited by the extensive computation necessary to determine the structure of a single grain boundary. If the structure could be predicted with more efficient computation, the understanding of the grain boundary would be accelerated significantly. Here, we predict grain boundary structures and energies using a machine-learning technique. Training data for non-linear regression of four symmetric-tilt grain boundaries of copper were used. The results of the regression analysis were used to predict 12 other grain boundary structures. The method accurately predicts both the structures and energies of grain boundaries. The method presented in this study is very general and can be utilized in understanding many complex interfaces.

**Introduction**

Polycrystalline materials comprised numerous crystal grains, and grain boundaries are formed by the attachment of different crystals. Since two differently oriented crystals form a grain boundary, the atomic arrangement at the grain boundary differs significantly from that in the bulk. This different atomic arrangement endows the grain boundary with peculiar properties, such as fast ion transportation and preferential deformation[1–7]. The atomic arrangement at the grain boundary is thus of significant interest to the research of materials science and engineering.

Computations of grain boundary structures using first-principles calculations and static lattice models have been extensively performed[8–10]. However, exhausting calculations are necessary to determine the complete grain boundary structure. This mainly results from the geometrical freedom of the grain boundary. Generally speaking, nine degrees of freedom, including five macroscopic and four microscopic, are present in a grain boundary[11]. Even in the coincidence site lattice (CSL) grain boundary, called the Σ grain boundary, four degrees of freedom of three-dimensional rigid body translations (Fig.1(a)) and one grain-boundary plane must be considered. With simplified CSL grain boundaries, the number of configurations to be considered still remains anywhere from 100 to 10,000, depending on the complexity of the grain boundary structure[12,13]. The theoretical calculations (static-lattice calculation or first-principles calculation) must be run by the number of these configurations to optimize the atomic structure and calculate the interface energy. Figure 1(b) shows the four dimensional plot of the three dimensional rigid body translation and the grain boundary energy. One has to determine the most stable point in this data set. Only after this computation can the most stable atomic configuration and its grain boundary energy for one grain boundary type be determined. This computation for determining grain boundary structures is truly exhaustive, limiting systematic studies of grain boundary interfaces to simple metal systems[14–16]. Despite the importance of these interfaces, accelerating the computation is not easy. For instance, Chua et al. reported predictions of the stable structure of complex-oxide grain boundaries using a generic algorithms[17–20]. Similar methods for predicting unknown grain boundaries were performed by *ab initio* random structure searching algorithms by Schustreritsch et al.[19] These methods have successfully predicted unknown grain boundary structures. However, even with these methods, many trial calculations are necessary to determine a single grain boundary structure. If the prediction of an unknown grain boundary can be achieved with less calculation, systematic investigations of more grain boundary types can be achieved within practical time frames. This systematic study of the grain boundary can provide deeper understanding of the structure-property relationships at this region.

Here, we demonstrate the prediction of the atomic arrangement and energy with the aid of a machine-learning technique. After training the relationships between the grain boundary energy and the interface structure, our method is very powerful in systematic investigations of the grain boundary.

**Results**

In this study, we focused on a series of [001]-axis symmetric-tilt CSL grain boundaries of face-centered copper, because many experimental and computational studies have been reported for this system to confirm our predictions. Since the CSL grain boundary of single-element materials are under consideration, the grain boundary possesses three degrees of freedom, namely in the rigid-body translation of one side of the crystal with respect to the other side of the crystal in three dimensions.

In this study, 17 CSL grain boundary types of Cu were considered: Σ5[001]/(210), Σ5[001]/(310), Σ13[001]/(230), Σ17[001]/(410), Σ17[001]/(350), Σ25[001]/(430), Σ25[001]/(710), Σ29[001]/(520), Σ29[001]/(730), Σ37[001]/(610), Σ37[001]/(750), Σ41[001]/(910), Σ41[001]/(540), Σ53[001]/(720), Σ53[001]/(950), Σ61[001]/(11 1 0), and Σ125[001]/(11 2 0). To obtain stable structures for these grain boundaries, we must consider approximately 1,000,000 configurations.

Here, Σ5[001]/(210), Σ5[001]/(310), Σ17[001]/(410) and Σ17[001]/(350) were selected as training data. For this training data, 150,000 configurations, or approximately 15% of the possible configurations, were used.

We performed regression analysis on the calculated grain boundary energy. To predict the grain boundary energy efficiently, selecting descriptors for the regression analysis is important. In this study, we obtained the descriptors from the geometrical data in "the initial atomic configuration," before theoretical calculation. This enables the prediction of the grain boundary energy, obtained after the calculation, without theoretical calculations. We can predict the most stable grain boundary without any calculation with the regression analysis results. The non-linear support-vector machine method was used in this study. The descriptors for the regression analysis are listed in the Supplementary Materials Table S1. Twelve geometric descriptors, such as atomic density and shortest and longest bond length, were obtained from the initial atomic configuration, and the squares, inverses, and exponentials of these data were also used.

In the support-vector regression analysis, three parameters of the epsilon-tube radius, cost, and variance of kernel function were determined, considering cross variation score and generalization ability. The result of the regression analysis is shown in Fig. 2(a). Most data lie on the gray line, indicating that the predicted energies are equal to the experimental energies. This demonstrates that the constructed prediction model has the ability to accurately describe the grain boundary energy using the provided descriptors.

To evaluate the accuracy of the prediction model, $\Sigma13[001]/(230)$ was used as test data. The results predicted by the model are shown in Fig. 2(b). Although the $\Sigma13[001]/(230)$ grain boundary was not used for the prediction model, the predicted grain boundary energy agrees well with the experimental grain boundary energy (Fig.2(b)). The most stable grain boundary configuration was predicted to be obtained at the rigid-body translations of X = 5.0 Å/13 Å, Y = 1.0 Å, and Z = 0.0 Å/3.6 Å. This allowed a one-time calculation for the rigid-body translation state. The calculated grain boundary energy with the predicted rigid-body translation state was 0.84 J/m$^2$, equivalent to the experimental grain boundary energy. The predicted stable and real structures are compared in Fig. 2(c)-(d). Both grain boundaries are equal, and they both contain periodic arrays of the 6-membered structure unit. This demonstrates that the method successfully predicts the candidate configuration providing a stable grain boundary; both grain boundary energy and atomic configuration are obtained by a one-time calculation of the predicted configuration. Notably, the predicted grain boundary energy before the one-time theoretical calculation was 0.96 J/m$^2$. Prior to any calculation, our method can predict the grain boundary energy within 10% error. The accuracy of the prediction is ascribed to the correlation between the initial atomic configuration and the grain boundary stability. This implies that our method is generally applicable to grain boundaries in any material system.

Here, this method is applied to another 12 grain boundaries of Σ25[001]/(430), Σ25[001]/(710), Σ29[001]/(520), Σ29[001]/(730), Σ37[001]/(610), Σ37[001]/(750), Σ41[001]/(910), Σ41[001]/(540), Σ53[001]/(720), Σ53[001]/(950), Σ61[001]/(11 1 0), and Σ125[001]/(11 2 0). As demonstrated with the test data, the candidate configuration providing the most stable structure was predicted by the model, and the grain boundary energy and atomic configuration were obtained by calculating this predicted configuration. Figure 3(a) shows the results of the predicted grain boundary energy and a comparison with previously reported grain boundary energies[14,21]. From the previous studies, the grain boundary energy shows a convex profile in relation to the misorientation angel. The energy gradually increases with increasing misorientation angle, reaching ~1.0 J/m$^2$ when the misorientation angle is 45°, and the energy decreases again at much higher misorientation angles. A detailed inspection reveals small cusps, or energy drops, at 28.07, 36.87, 53.13, and 67.38°, corresponding to Σ17[001]/(410), Σ5[001]/(310), Σ5[001]/(210), and Σ13[001]/(230), respectively. Fig. 3(a) presents the predicted grain boundary energies of all boundaries investigated in this study. The overall profile of the grain boundary energy, having a convex shape with a maximum at 45°, agrees well with the previous report. Small cusps at 36.87 and 53.13° are reproduced by the prediction model (cusps Σ5[001]/(310) and Σ5[001]/(210) are used in the training data). All atomic structures of the predicted grain boundaries are shown in Supplementary Materials Fig. S2, and the Σ53[001]/(720) and Σ41[001]/(540) are shown in Fig. 3(b)-(c). No reports exist on the atomic structure of these grain boundaries, but we expect that they are stable structures because their energies match the previously reported values.

Finally, we consider the effectiveness of our method. The method requires only one calculation for each boundary, while conventional methods require 1,000,000 calculations in total. Since the static-lattice potential method is used in this study, approximately 15 min are necessary for one calculation. If all calculations were performed in series as a single task, this would take 27 years to complete. However, once we achieve training, only 4 h are needed to calculate all grain boundary structures and energies. Our method increases efficiency by 58,000 times.

In summary, we attempted to predict the structures and energies of grain boundaries with a machine-learning technique. Geometrical factors before calculation, such as the shortest bond length and local atomic density, were selected as descriptors, and regression analysis was performed for the grain boundary energy using the non-linear supporting vector method. The grain boundary energies and structures of four grain boundary types, characterized by these descriptors, were used for training; the structures and energies of 12 other grain boundaries were successfully predicted. Although the static lattice method was used in this study because a simple metallic system was chosen, the first-principles method is mandatory to investigate the grain boundaries of complex compounds. In such cases, more exhaustive computation is necessary; however, the method described here is effective for such cases. Furthermore, the descriptors used in this study are geometrical factors, like bond length and atom density, measured before the calculation; the grain boundary energy can be described in terms of these descriptors. The present method has potential use in understanding complex grain boundaries, such as much higher $\Sigma$ and random grain boundaries. We believe that our method may permit the comprehensive understanding of grain boundary phenomena in many materials.

**Author Contributions**

T.M.(Mizoguchi) supervised this research, directed the calculations and wrote the paper. S.K. performed theoretical calculation, data analysis, and wrote the manuscript. T.M. (Miyata) discussed the results of this manuscript.

**Competing financial interests:** The authors declare no competing financial interests.

*Correspondence and requests for materials should be addressed to teru@iis.u-tokyo.ac.jp.

**Methodology**

With conventional methods, determination of stable structures requires the calculation of various configurations in which one grain has been translated in three directions relative to the basis position, such as mirror symmetry. In this study, the same method was used to construct a data space for regression analysis. The lattice static calculations were performed with the Gulp code[22]. Consider the *x* and *z* axes as vectors vertical to the grain boundary and the *y* axis a vector horizontal to the [001] direction. Rigid-body translations in the *x* and *z* directions had translational step sizes of 0.1 Å. Those in the *y* direction had step sizes ranging from 1.0 to 1.5 Å in increments of 0.1Å. To prevent the grain boundary structures from transforming to the bulk structure, atoms located furthest from the grain boundaries were fixed. As a result, 200,000 initial configurations were calculated for structural relaxation. The embedded-atom method potentials were used as empirical potentials[23]. The grain boundary energies were estimated by the following formula:

$$E_{GB} = \frac{E_{tot} - E_{bulk}}{2A}$$

where $E_{tot}$ is the total energy of the supercell with grain boundaries, $E_{bulk}$ is the total energy of the supercell without grain boundaries, and *A* is the grain boundary area.

**Support Vector Regression**

Support vector regression (SVR) is a non-linear regression analysis type based on the support vector machine (SVM), which is a discriminant function using the kernel function. SVR has the advantages of robustness and flexibility in alignment, avoiding unlike overfitting. In this study, to construct the prediction model, the most stable structures and metastable structures of Σ5[001]/(210), Σ5[001]/(310), Σ17[001]/(410), and Σ17[001]/(350) were considered. Using the Gaussian curve as the kernel function, SVR was performed with required parameters set as follows: 0.01 margin of tolerance, 1000 penalty factor, and 0.0001 variance. This regression analysis focused on the relationships between the grain boundary energy and the initial atomic configuration before calculation. Table 1 shows the descriptors used in performing SVR. In addition to these descriptors, their squares, inverses, exponentials, and exponential inverses were considered. In the result, 83 descriptors were standardized to align their average and variance to zero and one, respectively.

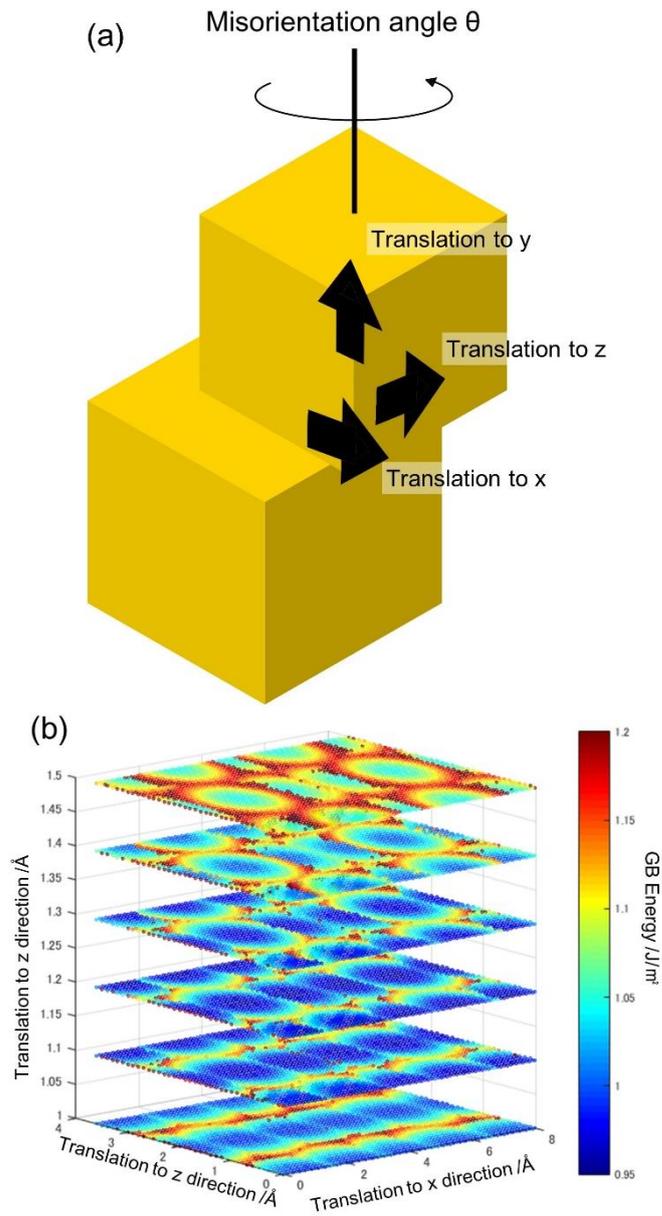

Figure 1 (a) Schematics of degrees of freedom for symmetric-tilt grain boundary. (b) Translation states‑grain boundary energy plot for $\Sigma 5[001]/(210)$.

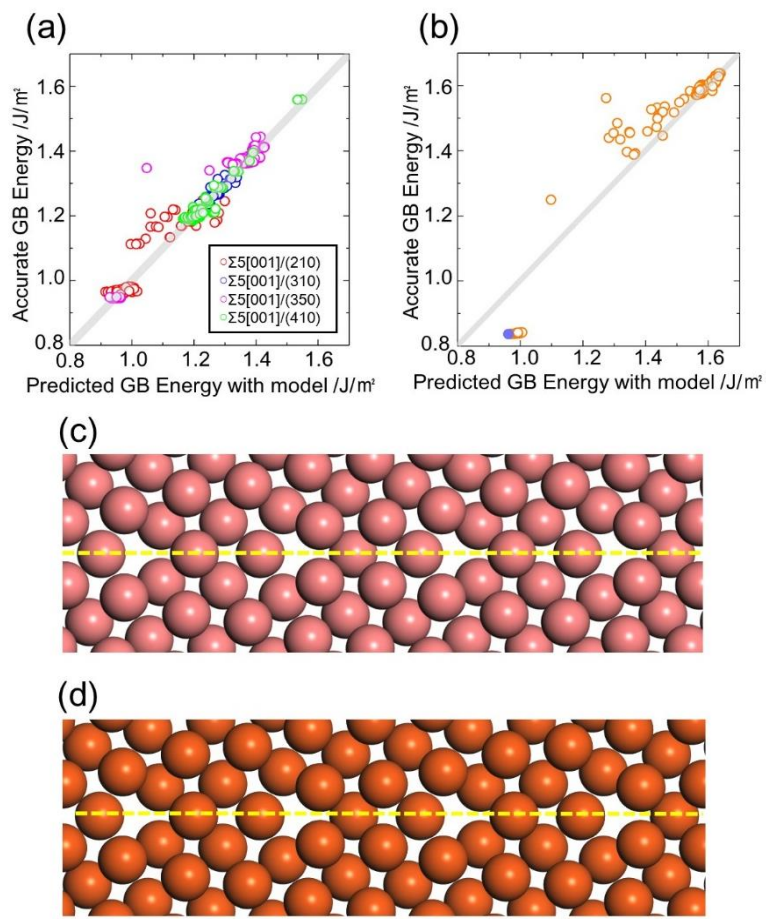

Figure 2 Predicted grain boundary energies and experimental grain boundary energies from the training data (a) and the test data (b). (c) The most stable structure of Σ13[001]/(230) and (d) the predicted structure by the present method. Yellow dashed lines represent the position of the grain boundary.

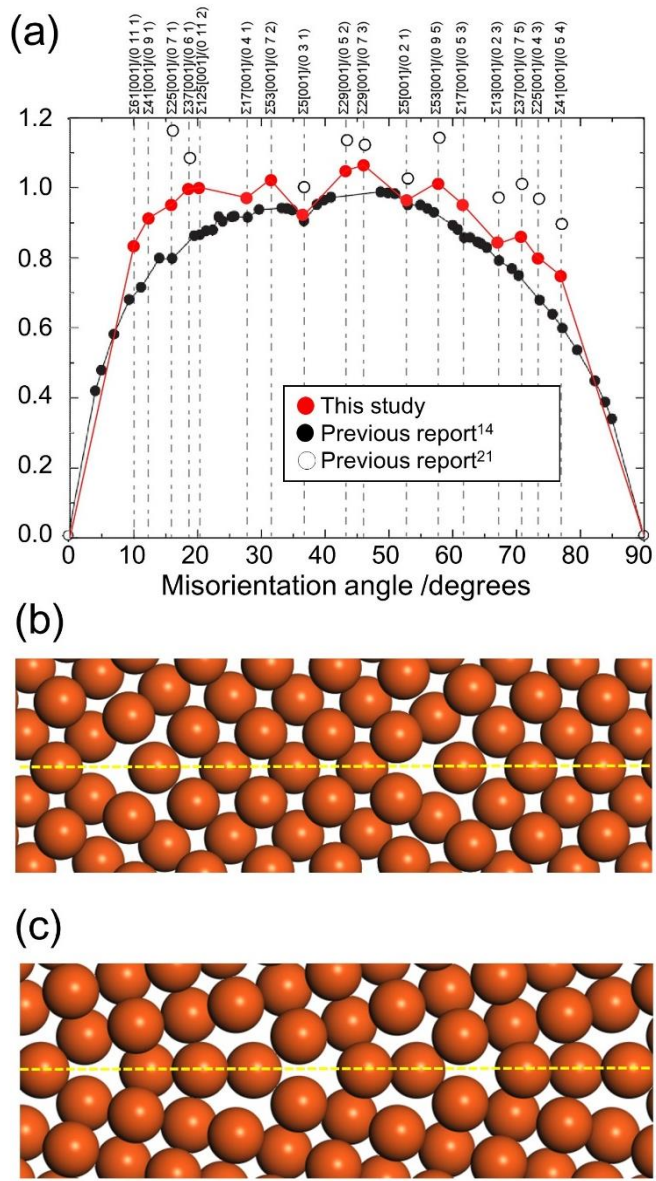

Figure.3 (a) Grain boundary energies are plotted with respect to misorientation angle. Red circles are obtained by the present method and open and filled black circles are obtained from the previous studies. (b)-(c) The predicted stable structures of the Σ41[001]/(540) and Σ53[001]/(720) boundaries.